\documentclass[
 reprint,
 amsmath,amssymb,
 aps,
prb,
]{revtex4-2}

\usepackage{subfiles}
\usepackage{graphicx}
\usepackage{dcolumn}
\usepackage{bm}
\usepackage{hyperref}
\usepackage[mathlines]{lineno}
\usepackage{subcaption}
\begin{document}
\preprint{APS/123-QED}

\title{High-resolution Measurements of Thermal Conductivity Matrix and Search for Thermal Hall Effect in La$_2$CuO$_4$}

\author{Jiayi Hu\textsuperscript{1}}
\author{Haozhi Xu\textsuperscript{1}}
\author{Juntao Yao\textsuperscript{2,3}}
\author{Genda Gu\textsuperscript{2}}
\author{Qiang Li\textsuperscript{2,4}}
\author{N. P. Ong\textsuperscript{1}}

\affiliation{{\normalfont\textsuperscript{1}}Department of Physics, Princeton University, Princeton, NJ, 08544, USA\\
    {\normalfont\textsuperscript{2}}Condensed Matter Physics and Materials Science Department, Brookhaven National Laboratory, Upton, NY, 11973, USA\\
    {\normalfont\textsuperscript{3}}Department of Materials Science and Chemical Engineering, Stony Brook University, Stony Brook, NY, 11794, USA\\
    {\normalfont\textsuperscript{4}}Department of Physics and Astronomy, Stony Brook University, Stony Brook, NY, 11794, USA\\
}

\date{\today}

\begin{abstract}
    We investigated the longitudinal thermal conductivity $\kappa_{xx}$ and thermal hall conductivity $\kappa_{xy}$ in La$_2$CuO$_4$ at temperatures $T$ between 2 and 20 K in magnetic fields $H$ up to 10 T. Within the temperature and field intervals studied, we do not resolve any thermal Hall signal with a conservative upper bound of $|\kappa_{xy}/T| <1\times10^{-4}$ ${\rm Wm^{-1}K^{-2}}$. The longitudinal thermal conductivity $\kappa_{xx}/T$ agrees well with previous studies, in both magnitude and $T$ dependence. In both channels, we performed measurements using the field-sweep protocol. To achieve high resolution, we carefully took into account relaxation effects after each step-increase in $H$. At low $T$, we find a linear decrease in $\kappa/T$ vs. $H$, as well as weak hysteresis near the meta-magnetic transition of the spin degrees.
    
\end{abstract}

\maketitle

\section{Introduction}

Despite nearly four decades of research, the cuprates continue to attract strong interest. Aside from the anomalously high critical temperature for superconductivity, the cuprates display a host of unusual normal-state properties arising from strong correlation of the mobile electrons as well as co-existence of competing orders over a broad interval of temperature~\cite{keimer2015quantum}. Recently, interest in the thermal Hall effect has also increased. In strongly correlated insulators, the thermal Hall conductivity $\kappa_{xy}$ provides an incisive probe charge-neutral excitations that may have exotic origins~\cite{katsura2010theory,lee2015thermal,yokoi2021half,czajka2023planar,matsumoto2011theoretical,chern2021sign}. Reports of sizeable thermal Hall signals in nominally insulating parent cuprates have attracted much interest. In these reports, the in-plane and out-of-plane $\kappa_{xy}$ are closely similar in magnitude. The isotropy has prompted the interpretation that $\kappa_{xy}$ arises from chiral phonons in the pseudogap phase of the cuprates~\cite{chiralPhonons,giantTHE}. By contrast, some theories propose an extrinsic effect arising from skew scattering of phonons~\cite{sun2022large,guo2022resonant,flebus2022charged}.

Here, we report high-resolution measurements of both $\kappa_{xy}$ and the longitudinal thermal conductivity $\kappa_{xx}$ in the parent cuprate, La$_2$CuO$_4$ (LCO), which was reported to exhibit some of the largest thermal Hall signals among the cuprates. While our $\kappa_{xx}$ agrees very well with published results (in its dependences on $T$ and $H$), we did not observe any thermal Hall conductivity signal in the interval of $T$ from 2 to 20 K in ${\bf H}$ up to 10 T. The results place an upper bound on $\kappa_{xy}$ that is 10-30$\times$ lower than signals in previous reports.

\section{Methods}
\label{sec:methods}

The sample is mounted on a brass stage which serves as the heat reservoir in a magnetic field ${\bf H}\parallel {\bf\hat z}$ applied normal to the CuO$_2$ plane ($\bf\hat z$ is parallel to the crystal's $c$-axis in Bmab notation). The heat current density ${\cal\vec{J}}_Q$ flows within the plane along the axis $\bf\hat x$.
With these axes, the in-plane longitudinal thermal conductivity is denoted by $\kappa_{xx}$ while the thermal Hall conductivity is $\kappa_{xy}$. We measured as-grown crystals of La$_2$CuO$_4$ with $T\rm{_N}\sim$ 255 K.

Protocols for thermal Hall experiments are of two types. In the field-swept (FS) protocol, $H$ is slowly incremented with $T$ rigorously regulated. In temperature-scan (TS) protocols, the crystal is first cooled with $\bf H$ fixed $\parallel{\bf\hat z}$ (in persistent mode), followed by a second cool down with $\bf H$ reversed. While the two protocols should give similar results when the thermal Hall angle is sizeable ($\kappa_{xy}/\kappa_{xx}>0.1$), we find that the FS protocol is more amenable to experimental control in high-resolution measurements of very small $\kappa_{xy}$ signals. The major artifactual contributions to the measured thermal Hall signal are (1) relaxation and (2) magnetocaloric effects following external perturbations (see below). In addition, hysteretic contributions, either intrinsic from magnetic order or extrinsic from slight misalignments of sensors should be eliminated. Using the FS protocol, we can resolve changes in $T$ of $\pm 0.1$ mK at 2.3 K and changes in $\kappa_{xy}/T$ of $4\times 10^{-5}$ $\rm Wm^{-1}K^{-2}$. 

\begin{figure}
    \centering
    \includegraphics[width=0.45\textwidth]{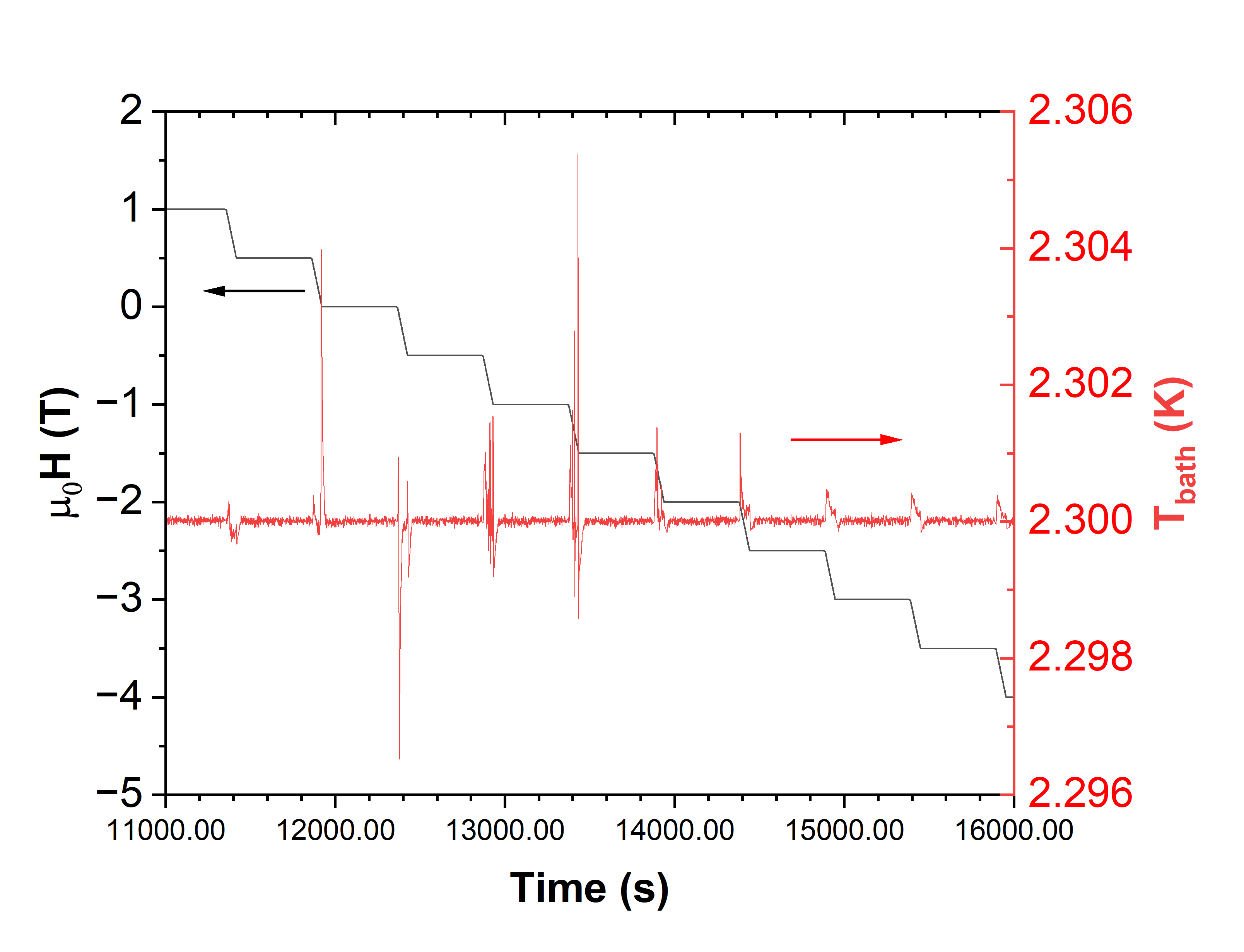}
    \caption{\label{fig_steps}Bath temperature relaxation captured during part of a stepped field sweep. The black line shows the applied magnetic field and the red line indicates the bath temperature. This particular sweep was taken with the heat bath set to 2.3 K and the field swept from 10 T to -10 T with a step-size of 0.5 T. As shown, a field step induces a temperature spike, attributed to the magnetocaloric effect and eddy currents in the metallic parts of the reservoir. The spikes relax on a time scale of 1-10 ms. After each step, we wait for 7 minutes before averaging the data within an acquisition window. A typical set of FS measurements at a fixed temperature consists of four sweeps, as described in Section \ref{subsec: FS}. As shown, our resolution in $T$ at 2.3 K is $\pm 0.1$ mK.
    }
\end{figure}

\subsection{Field-swept experiment}
\label{subsec: FS}
Aside from minor changes, we adopted the FS protocols described in Refs.~\cite{MaxThesis,PeterThesis}. With $T$ actively regulated at a selected value, we step-change $H$ in steps of heights between 0.1 and 1 T every 500 s. Figure \ref{fig_steps} shows the staircase profile of $H$ vs. time $t$ (black curve) at $T$ = 2.3 K. Following each step-change, the bath temperature $T_{bath}$ exhibits spikes of size 1 to 6 mK despite active regulation of $T_{bath}$ (red curve). These spikes, lasting up to 10 s, arise from the competing effects of eddy current heating of the bath and magnetocaloric effects in the sample. To exclude these non-equilibrium contributions to the $\kappa_{xy}$ signal, we wait 3 to 8 min. before acquiring and digitizing the thermal Hall signal within a narrow window in $t$. We can resolve changes in $T$ to $\pm 100 \;\mu$K.

At each $T$ and $H$, three CX-1030 thermometers are employed to determine $T_{bath}$ as well as the two thermal gradients $\nabla T\parallel {\bf \hat x}$ and ${\bf \hat y}$. 
During each cool-down from 300 K, the thermometers are calibrated \textit{in-situ} against a primary thermometer. To account for the intrinsic magnetoresistance of each thermometer, we repeated the calibration sweeps using the same field-step increments but with ${\cal\vec{J}}_Q$ turned off. Additionally, the bath temperatures were set to the highest and lowest values of the measured temperatures found in that pair of field sweeps. Hence for each sample thermometer at each specific field step, we may estimate the leading-order correction under field by interpolating between its resistance at the higher and lower bounds of the temperature range measured. An example of data before and after the correction is shown in the Supplement~\cite{supplement}.

Consecutive sweep-up and sweep-down scans of $H$ with $T$ fixed provide an effective way to eliminate hysteretic artifacts in the thermal Hall signals, using the ``in-out" technique~\cite{MaxThesis,PeterThesis}. Here, the ``external" component may arise from long relaxation times of the environment or by contamination from the longitudinal conductivity due to very slight misalignment in the thermometers. We assumed a finite contribution (with the unknown ratio $\alpha$) in the raw measurements of the transverse gradients: $\Delta T_{xy}(H)=\alpha\,\delta T_{xx}(H)+\delta T_{xy}(H)$, where $\Delta T_{xy}$ is the measured signal and $\delta T_{xx}$ and $\delta T_{xy}$ are the intrinsic values in an ideal setup. We then extracted $\delta T_{xy}$ by subtracting the field sweeps taken in the opposite direction: $\delta T_{xy}(H)=\frac12(\Delta T_{xy}(H)-\Delta T^*_{xy} (-H))$, so that the longitudinal component cancels out due to $\delta T_{xx}(H)=\delta T^*_{xx}(-H)$, where $\delta T_{xx}^*$ denotes the longitudinal contribution to the thermal Hall measurements taken at the opposite field sweep. We do indeed observe a small hysteresis in the longitudinal $\kappa_{xx}$ channel. For details, see Section~\ref{subsec:hysteresis}.

\subsection{Temperature sweeps}
As mentioned, we have also performed TS experiments.
When performing temperature sweeps (from $T_{low}$ to $T_{high}$), we ramp up the field to say 10 T at $T_{high}$ (roughly 20 K, restricted by the operation range of the $^3$He insert). We then acquire data while decreasing (or increasing) $T$ in steps. After the system returns to  $T_{high}$ and is fully equilibrated, we reverse the field direction and repeat the measurements using the same step profiles in $T$. Finally, the two curves $\Delta T_{xx}$ and $\Delta T_{xy}$ vs. $T$ are symmetrized and anti-symmetrized with respect to $\bf H$ to calculate the corresponding thermal conductivities. In TS protocols, there does not appear to be a way to detect and eliminate the transient, non-equilibrium effects displayed in Fig. \ref{fig_steps}. We did not go through the ``in-out" approach here to stay consistent with the commonly employed procedures. We attribute the much larger signal fluctuations (compared with the FS protocol) to the long time interval between up-field and down-field measurements. This also precludes applying the cancellation analysis used in the FS protocol.

\section{Results}
\label{sec:results}

\subsection{Absence of Thermal Hall}

Figure \ref{figtemp} plots the temperature dependence of the longitudinal thermal conductivity $\kappa_{xx}/T$ and thermal hall $\kappa_{xy}/T$ (both divided by $T$) at temperatures down to 2 K at the field values indicated.
Figure \ref{figtemp}a exhibits the data obtained from field sweeps (FS) while Fig. \ref{figtemp}b shows Hall data from temperature sweeps (TS), as described in Section \ref{sec:methods}. Note that the Hall data in panels (a) and (b) are displayed in units of  $10^{-3}$W/mK$^2$, which is $1,000\times$ smaller than that used in the inset of Panel (c). 
As anticipated, the FS protocol yields data that are much less noisy compared with the TS protocol. Interestingly, as $T$ decreases to 2 K, fluctuations in the FS data decrease whereas they increase in the TS data. A direct comparison of data collected by the two methods can be found in the Supplement~\cite{supplement}. 

In either case, we could not resolve any thermal hall signal in magnetic fields up to 10 T, and at temperatures as low as 2 K. 
All transverse temperature differences $T_{xy}$ deviating from zero fall within our resolution threshold $10^{-5}\sim10^{-4}$ W/mK$^2$ (when converted to $\kappa_{xy}/T$. Our resolution threshold is smaller by more than an order-of-magnitude than the magnitudes in previous reports (approximately $3\times10^{-3}$ W/mK$^2$ near $T$ = 7 K in a field $H$ = 15 T)~\cite{chiralPhonons}.

The resolution bounds for the transverse gradient $\Delta T_{xy}$ are independent of $T$ from 2 to 10 K. This is not compatible with the exponential decay of $\kappa_{xy}/T$ vs. $T$ proposed in previous publications. 
In our experiment, fluctuations in the transverse thermal Hall signal is roughly 10$\times$ larger in the TS scans compared with the FS scans. The largest peaks in the fluctuations are comparable to the signal $\kappa_{xy}/T$ signal in previous reports ($1.2\times10^{-3}$ W/mK$^2$). However, in our case, these are fluctuations that change sign randomly. When we perform FS scans, they are strongly suppressed or absent altogether.

The error bars in Fig.\ref{figtemp}c are estimated based on percentage uncertainties in both $\Delta T_{xx}$ and $\Delta T_{xy}$. As expected, the contribution of the transverse channel dominates that of the longitudinal. For example, at 16.5 K with $H$ = 10 T, the longitudinal signals contribute 0.6\% to the percentage uncertainties, while the transverse contributions reach 73\%. As shown, the increase in the error bars with $T$ is primarily caused by fluctuations in $\kappa_{xy}/T=0$.

Aside from factors such as sensitivity of the thermometers and the vacuum quality, the resolution of thermal Hall measurements is especially sensitive to the regulation (or lack thereof) of the bath temperature. We attribute the high resolution in ($\delta T_{xy} \simeq 30\,\mu$K at 2.3 K which translates to $4\times 10^{-5}$ W/mK$^2$) to the unusually large reservoir volume in our $^3$He insert, and the high signal-to-noise resolution in the FS protocol resolution. At higher temperatures, the stability of the Helium-3 insert is poorer because of the reduced cooling power. Our resolution estimates at 15.6 K is $2\,m$K, or $1.6\times10^{-4}$ $\rm Wm^{-1}K^2$ in $\kappa_{xy}/T$, which is an order of magnitude smaller than the thermal Hall value reported in the literature.
We conclude that the absence of resolvable thermal Hall signal over the temperature interval (2 to 20 K) is intrinsic and not attributable to artifactual constraints.

\subsection{Longitudinal thermal conductivity at Low Temperature}
We turn next to the longitudinal thermal conductivity $\kappa_{xx}$. Unlike the thermal Hall channel, our longitudinal thermal conductivity data nominally agree with previous reports aside from a weak anomaly at 7 K. Parts of the discussion are left to Sections \ref{subsec:field}.

\begin{figure*}
    \centering
    \includegraphics[width=\textwidth]{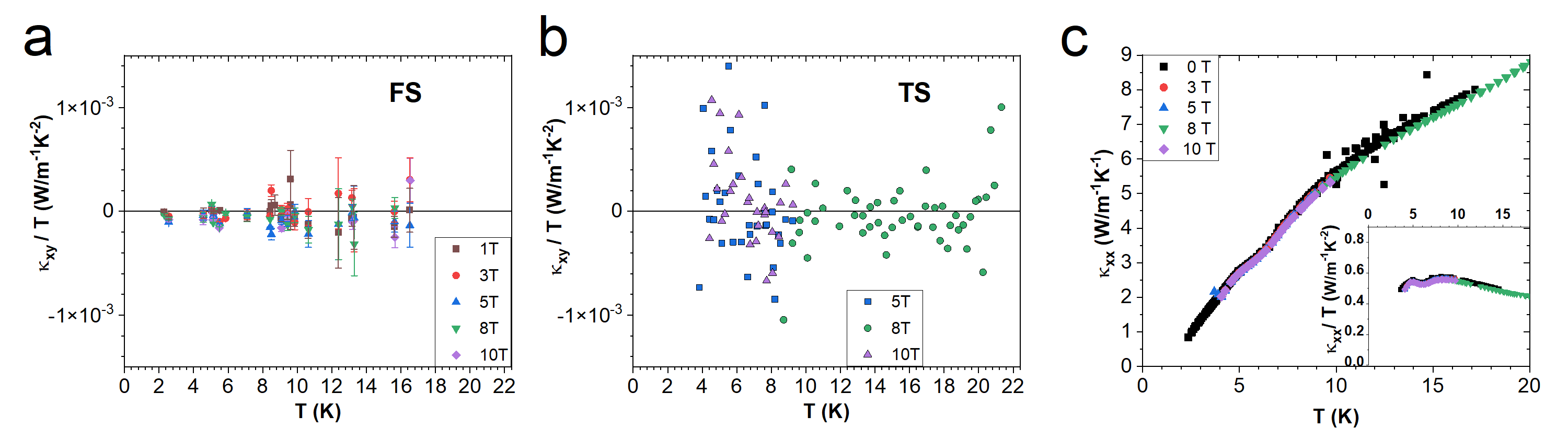}
    \caption{\label{figtemp}Temperature dependence of thermal Hall conductivity and longitudinal thermal conductivity measured in La$_2$CuO$_4$. Different colors correspond to different magnitudes of the applied magnetic field. Panels (a) and (b) plot $\kappa_{xy}/T$ versus temperature collected via field sweep (FS) and temperature sweep (TS) methods as mentioned in Section\ref{sec:methods}. The solid line $\kappa_{xy}/T=0$ is also plotted for reference. As demonstrated, despite that TS result in larger noise level, both sets of data suggest absence of thermal Hall. Panel (c) plots the temperature profile of the longitudinal thermal conductivity $\kappa_{xx}$, and the inset shows $\kappa_{xx}/T$ with a plot scale 1000 times larger than the other panels. We did not observe noticeable difference between TS and FS here.}
\end{figure*}

Figure \ref{figtemp}c plots the temperature dependence of the longitudinal thermal conductivity (magnitude of $H$ given by the color scale). To facilitate comparison with the thermal Hall data, we also show $\kappa_{xx}/T$ in the inset. As we cool from 20 to 2 K, $\kappa_{xx}$ decreases monotonically. A shoulder-like feature is resolvable near 7 K. As shown in the inset, this 6-K feature appears as a sharp dip followed by a weak peak (dividing by $T$ enhances the peak at 5 K). At present the 6-K is not understood despite many tests to rule out artifacts. Increasing $H$ from 0 to 10 T does not have any observable effect (Fig.\ref{figtemp}c) aside from a slight suppression of the peak in high fields. We defer discussion of the effect of $H$ to the next section.

In contrast to the thermal Hall data, the $\kappa_{xx}$ data are consistent with previous reports in both magnitudes and $T$ dependence~\cite{chiralPhonons,hess2003magnon}. While the previous works did not cool down to our lowest $T$ (2.5 K), our measurements near 20 K gives $\kappa_{xx}=8.7$ W/mK, in nomincal agreement with $\sim7$ W/mK in Ref.~\cite{hess2003magnon} and $\sim10$ W/mK in Ref.~\cite{chiralPhonons} measured at 15 T. While previous reports did not include longitudinal conductivity data below 7 K, data in the overlapping interval ($7, 20$) K matches well with our results. Whereas $\kappa_{xx}$ decreases monotonically as $T$ decreases, the ratio $\kappa_{xx}/T$ exhibits a broad peak near 15 K. Whereas we apply a lower field (10 T compared to 15 T), the weak field-dependence justifies the comparison here. The very small values of the fluctuating signals for $\kappa_{xy}/T$ below 20 K is incompatible with an exponential dependence on $T$. See the Supplement~\cite{supplement} for a discussion of the fits.

\subsection{Field dependence}
\label{subsec:field}

\subsubsection{Weak suppression of $\kappa_{xx}$ at high fields}
In Fig.\ref{figfield}, we focus on the field dependence and present the data obtained via the FS protocol (with $T$ indicated by the color scale). 
The rather high N\'{e}el temperature ($T_N\sim255\rm{K}$) implies that the magnetic order is well established at 4 K. Hence $\kappa_{xx}$ should vary weakly with $H$. Indeed, the field profiles are rather flat (Fig.\ref{figtemp}).
At 16.7 K, a 10-T field leads to a decrease in $\kappa_{xx}$ of only $\sim$ 2\%. To amplify the field dependence, we show in panel (a) the fractional change $\Delta\kappa_{xx}/\kappa_0\equiv(\kappa_{xx}(H)-\kappa_0)/\kappa_0$, where $\kappa_0\equiv\kappa_{xx}(0\rm{T})$). (A plot of the unsubtracted $\kappa_{xx}$ is shown in the Supplement~\cite{supplement}).

Above 8 K, the decrease in $\kappa_{xx}$ with $H$ is monotonic and linear in $H$ up to our highest field (10 T). We attribute the suppression to an increase in the spin-phonon scattering, as a large $H$ weakens the long-range antiferromagnetic order parameter.
However, starting at 6.1 K, the curve flattens out above 8 T. As we cool further to 4.5 K, the flat region evolves into a distinct up-turn that onsets at 7.5 T. These interesting changes are not understood.

\subsubsection{Hysteresis from weak ferromagnetism}
\label{subsec:hysteresis}
\begin{figure}
    \centering
    \includegraphics[width=0.48\textwidth]{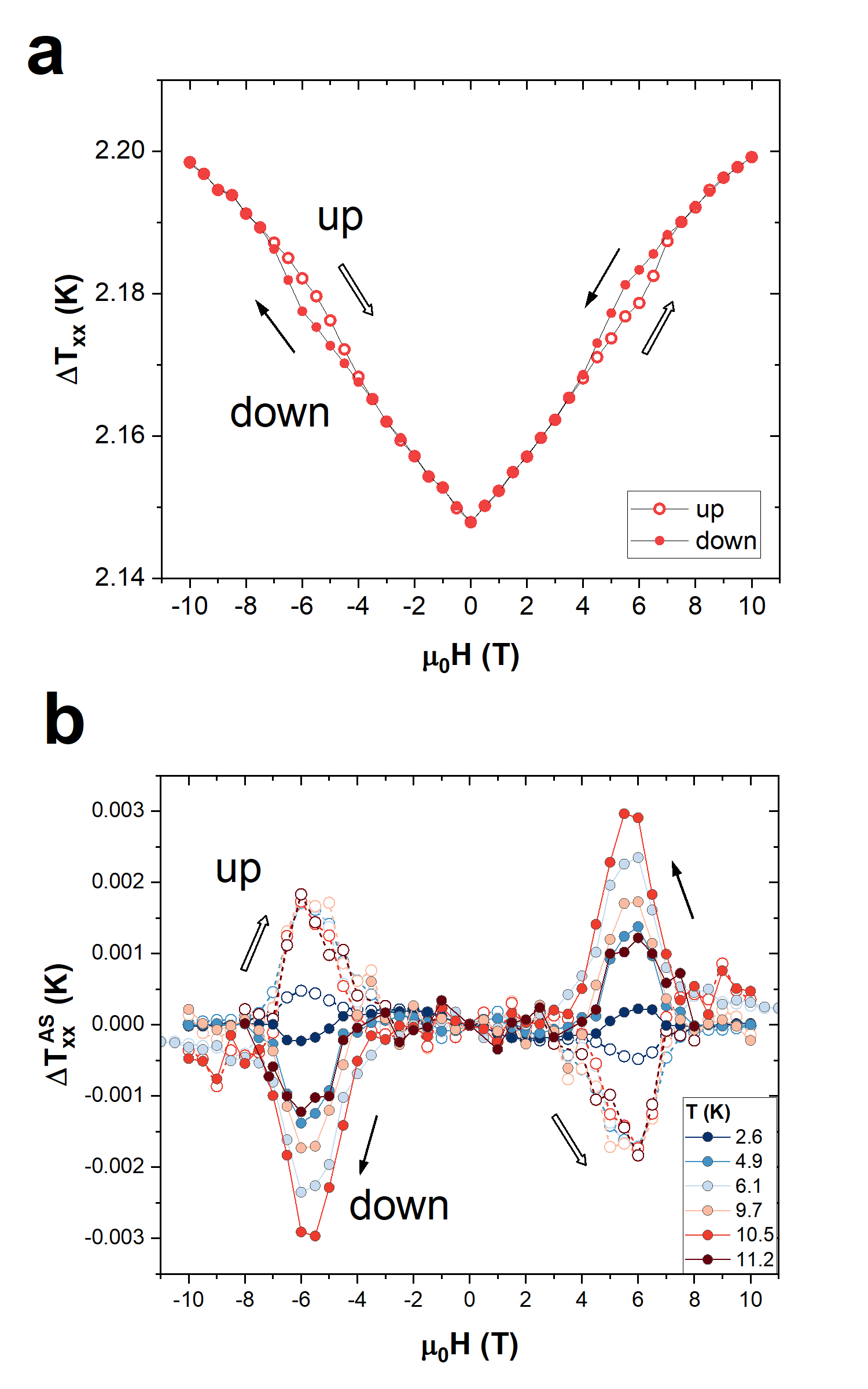}
    \caption{\label{hyst}Hysteresis loops observed in temperature gradients. Panel (a) shows the raw data of the longitudinal temperature gradients ($\Delta T_{xx}$) in a set of field sweeps measured at $10.52\pm0.02$ K. The solid circles are data points measured with the magnetic field swept from 10 T to -10 T, and the hollow ones with field in the opposite direction, swept upwards. Panel (b) plots examples of $\Delta T_{xx}$ after direct anti-symmetrization with respect to field, taken at different temperatures. The magnitudes of the transition fields stay near 4 T and 7 T.}
\end{figure}

In this section, we discuss the weak hysteretic features observed in raw data of both longitudinal and transverse temperature gradients, which comes from the small ferromagnetic moment previously analyzed in neutron scattering experiments~\cite{thio1988,thio1990,kastner1988neutron,Reehuis2006}. Figure \ref{hyst}a plots the change in $\Delta T_{xx}$ without any symmetrization in field scans at 10.5 K. Open and solid red circles correspond to up-field and down-field scans, respectively. Weak hysteretic loops appear in the field intervals $\pm(4,7)$ T.

We also note similar features in raw data of $\Delta T_{xy}$, but smaller in magnitudes and distorted in shape because of admixing with the longitudinal signal. In this case, a naive antisymmetrization in $\Delta T_{xy}$ could result in a misleading thermal Hall signals. Through direct antisymmetrizing $\Delta T_{xx}$, we amplify the hysteresis contribution and present some examples of data taken at different temperatures in Fig.\ref{hyst}b. While further study might be necessary to draw conclusions on how the peak heights of the hysteresis loops evolve as temperature changes, the critical field values stay rather consistent near $|H|=4\sim7$ T over the measured temperature range, with a slight narrowing at the high temperature tail.

The hysteretic features are consistent with meta-magnetic transitions inferred from measurements of resistance, magnetization, and neutron diffraction~\cite{thio1988,cheong1989,kastner1988neutron,hu2023spin}. In 1988, Thio et. al. report magnetic transition in electrical transport and susceptibility measurements. They explain its origin with the ferromagnetic moment arising from a small canting of moment out of the Cu-O plane, and hence applying a field $H//\hat{c}$ that's stronger than the interlayer antiferromagnetic coupling induces a spin-flop transition across the layers. With a recent neutron scattering experiment that suggests a specific pattern of spin rotation, the field-induced meta-magnetic transition into a weak ferromagnetic phase appears to be well supported by a number of experiments. The transition field has been reported to range from 5 to 11.5 Tesla below 10 K for samples with slightly different stoichiometry~\cite{kastner1988neutron,Reehuis2006,hu2023spin,supplement}.

\subsubsection{Upper bound for thermal Hall conductivity}
\begin{figure*}
    \centering
    \includegraphics[width=0.9\textwidth]{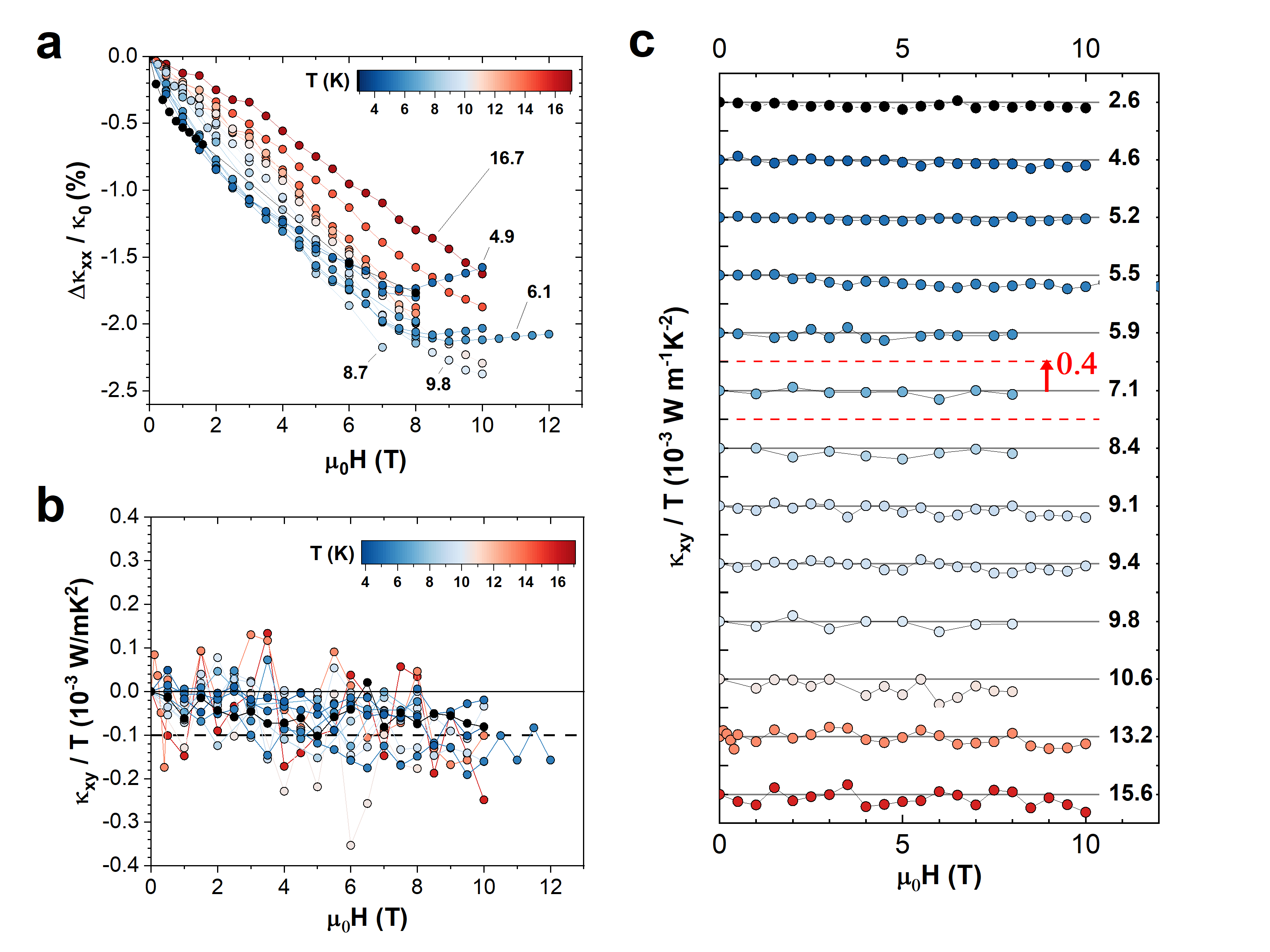}
    \caption{\label{figfield} Field dependence of thermal conductivities taken via FS method. Panel (a) plots the percentage change in $\kappa_{xx}$ vs. $H$ at different temperatures. In general, $\kappa_{xx}$ is suppressed by $H$. The decrease is approximately linear in $H$ (up to 10 T). Below 6.1 K, however, $\kappa_{xx}$ flattens to a plateau. At 4.9 K, $\kappa_{xx}$ begins to increase with $H$ above 8 T.
    Panel (b) plots in high resolution our measurements of the thermal Hall signal (expressed as $\kappa_{xy}/T$) at the temperatures $T$ indicated by the horizontal color-scale bar. The dashed line at $\kappa_{xy}/T=-1\times10^{-4}$ $\rm Wm^{-1}K^{-2}$ indicates the bound dictated by uncertainties in our experiment. Panel (c) displays the traces of the thermal Hall signal (again expressed as $\kappa_{xy}/T$). The curves are individually displaced for clarity. At each $T$ (shown on the right column), the thermal Hall signals fluctuate about the zero value represented by the solid lines. The vertical scale, identical for all curves, is given by the red arrow which extends from $0$ to $\to$ $0.4\times10^{-3}$ $\rm Wm^{-1}K^{-2}$.}
\end{figure*}

 In Fig. \ref{figfield}b we plot field profiles of the transverse thermal signal (expressed as $\kappa_{xy}/T$). The color scale indicates the temperature. The dashed line indicates our resolution. For clarity, we replot each of 13 curves vertically displaced in Fig. \ref{figfield}c with $T$ given on the right (at each $T$, the zero value is drawn as the solid line). For each curve, the vertical scale is $\pm 0.4\times 10^{-3}$ in units of ${\rm Wm^{-1}K^{-2}}$, as indicated by the arrows and the dashed lines sandwiching the curve at 7.1 K. We note again that the resolution of the thermal Hall data is $1000\times$ higher than that used in Fig.\ref{figtemp}c for $\kappa_{xx}/T$. 

If we average by eye the entire set of curves in Fig. \ref{figfield}b, it is plausible that there may exist a very weak thermal Hall signal of roughly $0.1\times 10^{-3}$ ${\rm Wm^{-1}K^{-2}}$. However, when the curves are individually displaced in Fig. \ref{figfield}c, this conclusion is seen to be invalid. The weak positive average is largely skewed by the 2 curves at 5.5 and 10.6 K. We may take the value $0.1\times 10^{-3}$ ${\rm Wm^{-1}K^{-2}}$ as a conservative upper bound for our thermal Hall measurement. 
This upper bound is 10 to 30 $\times$ (depending on $T$) smaller than the values reported in Ref.~\cite{chiralPhonons}.

\section{Discussion and Summary}

A plausible reason for the strong disagreement between our thermal Hall measurement and previous experiments may come from subtle sample differences. We investigated as-grown single crystals of the cuprate parent compound La$_2$CuO$_4$ with $T_N$ near 255 K, while samples measured in previous literature could have slightly different stoichiometry caused for e.g. by deoxygenation. Although the magnitude of the spin gap greatly exceeds $T$ at the temperatures studied here, features intrinsic to weak ferromagnetism sensitive to oxygen content may play a role. Nonetheless, our experiment raises concern on the interesting question whether an intrinsic thermal Hall effect exists in La$_2$CuO$_4$.

In summary, we have performed measurements of the longitudinal thermal conductivity in La$_2$CuO$_4$. We also performed a high-resolution  search for the thermal Hall signal between 2 and 20 K. Our results for the longitudinal thermal conductivity $\kappa_{xx}$ agree substantially with previous studies. This includes fine details like the weak hysteresis associated with meta-magnetism related to field-induced ordering of the moments. However, we did not observe a finite value in fields up to 10 T, in contrast to the report of a large negative thermal Hall effect that decays exponentially with $T$. Taking account of the main uncertainties and resolution limits, we may set a (conservative) upper bound on $|\kappa_{xy}/T|$ of $0.1\times10^{-3}$ $\rm Wm^{-1}K^{-2}$ in the interval $2<T<20$ K. We hope these results will motivate other groups to revisit this interesting problem.

\clearpage
\newpage

\bibliography{citations}

\clearpage
\newpage
\begin{center}
\section*{Supplemental Figures}

\renewcommand{\thefigure}{S\arabic{figure}}
\title{SUPPLEMENTAL MATERIAL\vspace{5mm}\\High-resolution Thermal Hall and Thermal Conductivity Measurements of Cuprate Parent Compound La$_2$CuO$_4$ at Cryogenic Temperature\\
\vspace{5mm}
Jiayi Hu, Haozhi Xu, Juntao Yao, Genda Gu, Qiang Li and N. P. Ong}

\maketitle

\begin{figure}[h]
    \centering
    \includegraphics[width=0.45\textwidth]{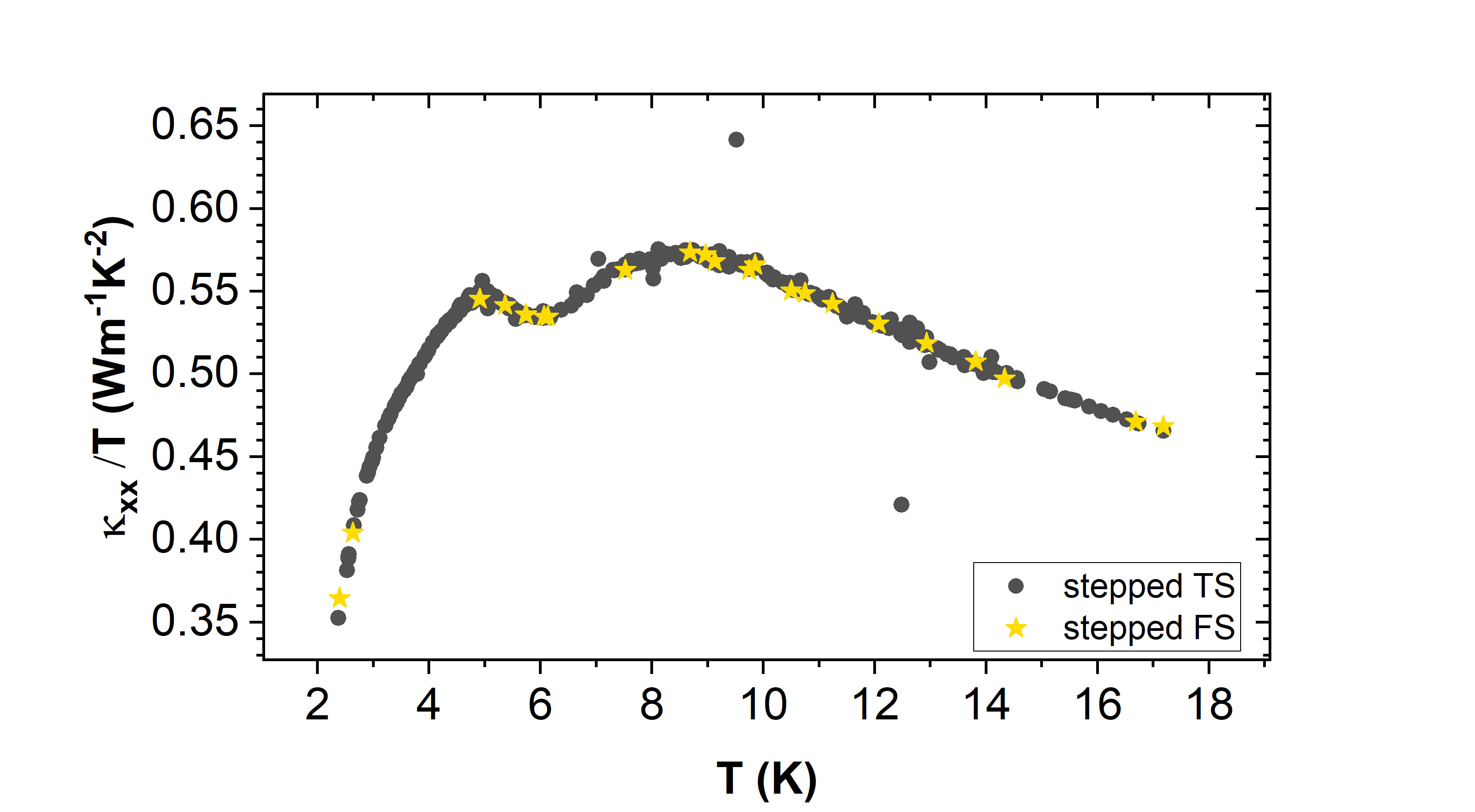}
    \caption{\label{kxx_TSFS} Comparing results of longitudinal thermal conductivity measured via temperature sweeps (TS) and field sweeps (FS) methods at zero field. The yellow stars are data collected when sweeping field in steps and the black squares are when from sweeping temperature. The apparent outliers are measurements taken with heat current too low for the specific temperatures. We do not observe any noticeable difference between employing the TS method and the FS method for $\kappa_{xx}$/T.}
\end{figure}

\begin{figure}[h]
    \centering
    \includegraphics[width=0.45\textwidth]{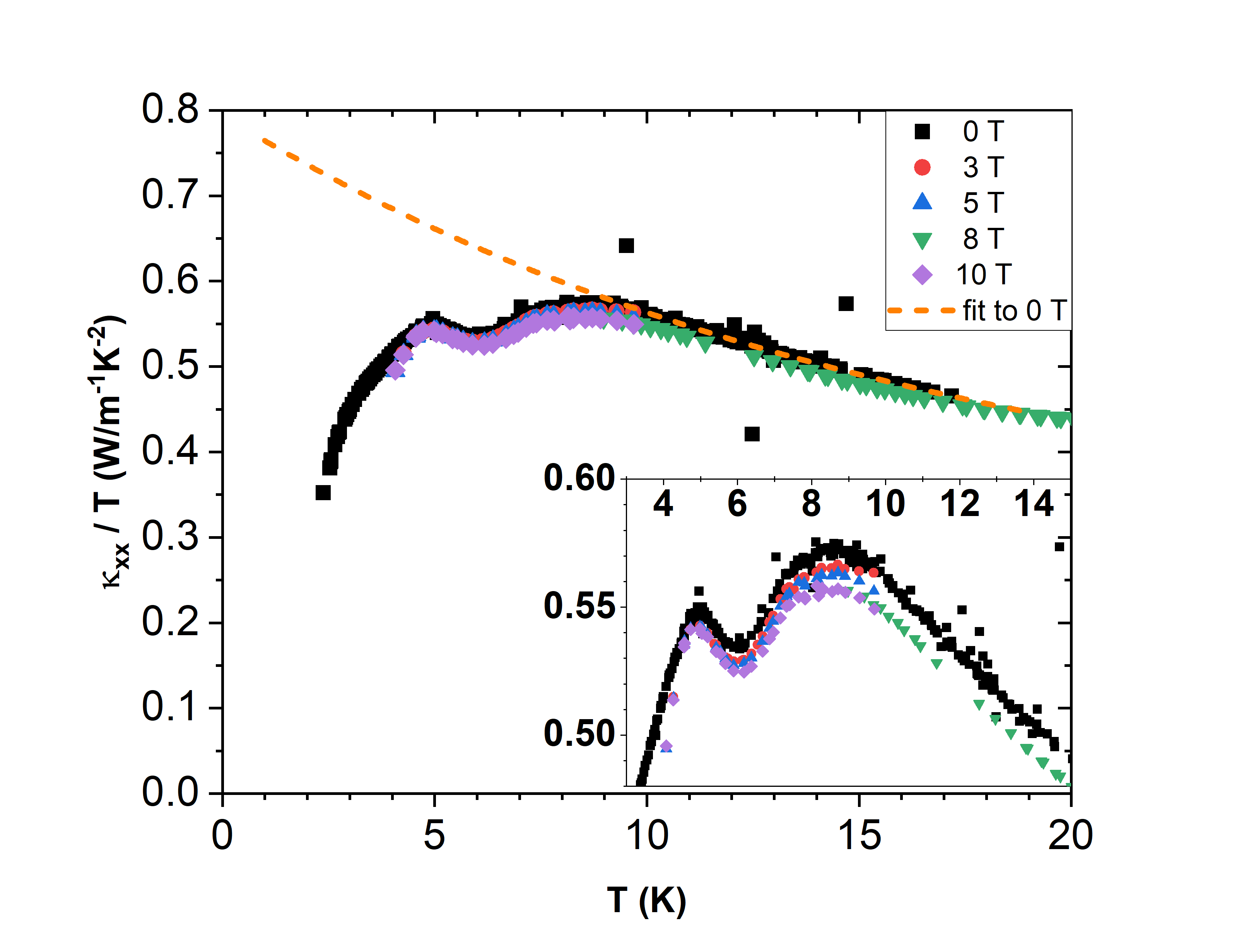}
    \caption{\label{kxx_exp} Zoom-in on the longitudinal thermal conductivity $\kappa_{xx}/T$. The orange dashed line shows a fit to the exponential $A\,\text{exp}(-T/T_0)+C$ with best-fit parameter $T_0=16.5$ K.}
\end{figure}

\begin{figure}
    \centering
    \includegraphics[width=0.45\textwidth]{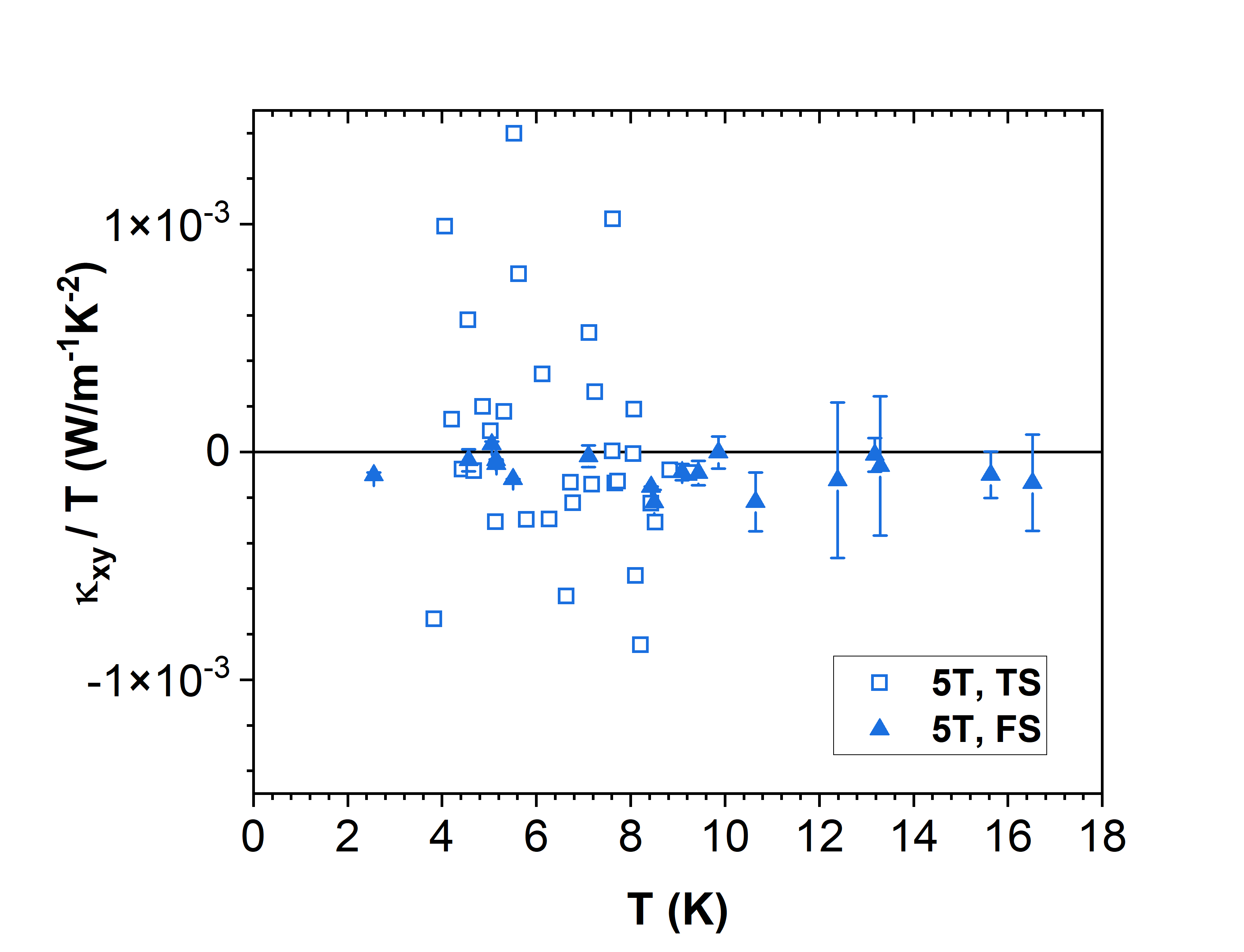}
    \caption{Comparing results of thermal Hall conductivity measured via TS and FS methods at 5 T. The hollow squares correspond to TS and the solid triangles corresponds to FS methods, with similar heat current through the sample.}
    \label{kxy_TSFS}
\end{figure}

\begin{figure}[h]
    \centering
    \includegraphics[width=0.45\textwidth]{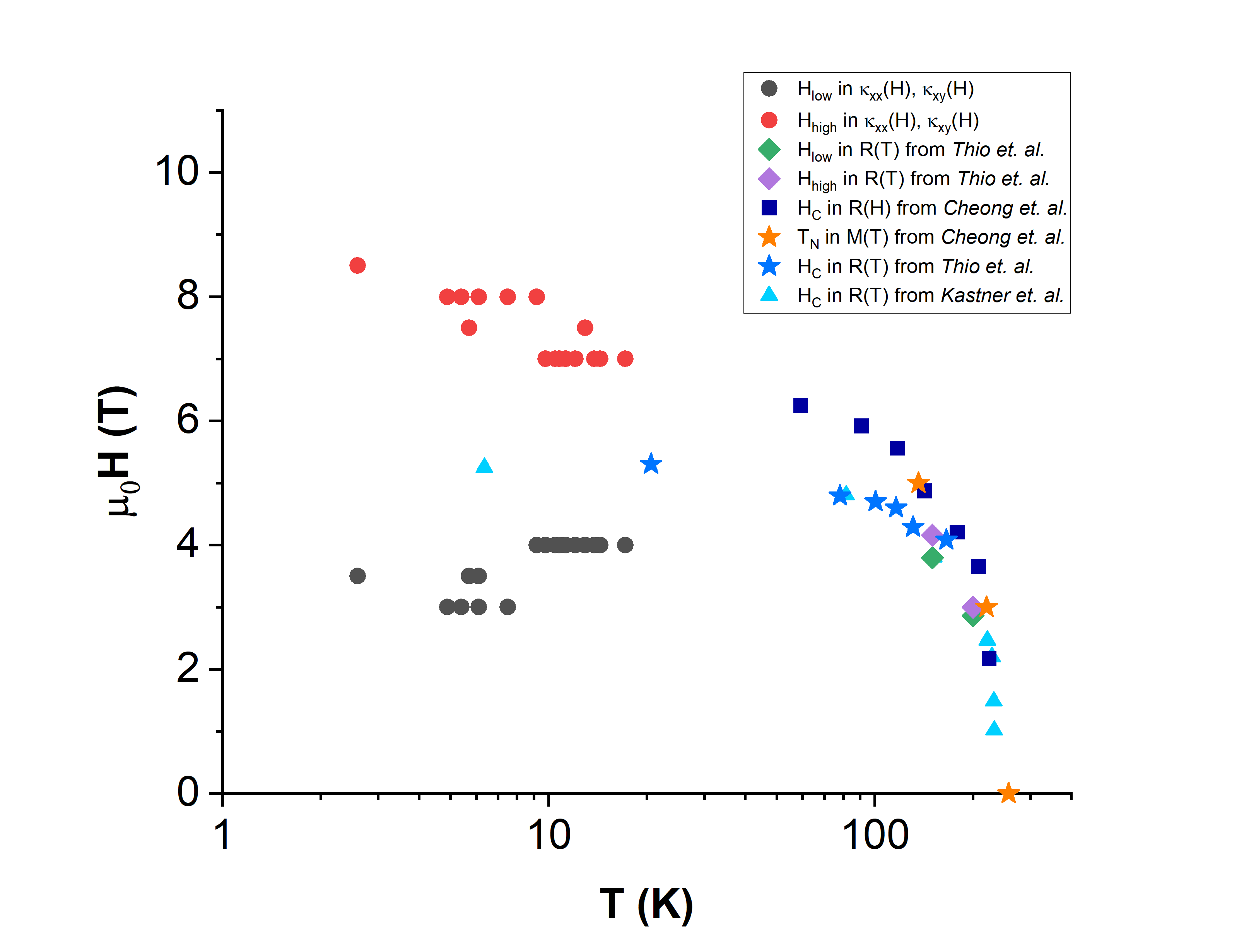}
    \caption{Critical field values of the hysteresis observed in thermal conductivities and in the literature. Red and black circles are the upper and lower boundaries of the hysteresis loops read out from $\Delta T_{xx}$ and $\Delta T_{xy}$ in this study. The other symbols are read out from electrical resistivity and magnetization measurements in the literature~\cite{thio1988,thio1990,cheong1988,cheong1989,kastner1988neutron}.}
    \label{fig:WFM}
\end{figure}


\end{center}

\end{document}